\documentclass[aps,prb,amsmath,amssymb,twocolumn,longbibliography, superscriptaddress]{revtex4-2}
\usepackage{graphicx}
\usepackage{dcolumn}
\usepackage{bm}
\usepackage{mathrsfs}
\usepackage{subfigure}
\usepackage{natbib}
\usepackage{amssymb}
\usepackage{multirow}
\usepackage{float}
\usepackage{color}
\usepackage{ulem}
\usepackage{mathrsfs}
\usepackage[colorlinks,linkcolor=blue,anchorcolor=blue,citecolor=blue,urlcolor=blue]{hyperref}

\begin{document}

\title{Superconducting properties of Fibonacci chains with enhanced superconducting pairing at the boundaries}

\author{Quanyong Zhu}
\affiliation{School of Mathematics and Computer Science, Lishui University, 323000 Lishui, China}

\author{Guo-Qiao Zha}
\affiliation{Physics Department, Shanghai Key Laboratory of High Temperature Superconductors, Shanghai University, Shanghai 200444, China}

\author{A. A. Shanenko}
\affiliation{HSE University, 101000 Moscow, Russia}
\affiliation{Moscow Center for Advanced Studies, Kulakova str. 20, Moscow 123592, Russia}

\author{Yajiang Chen}
\email{yjchen@zstu.edu.cn}
\affiliation{Zhejiang Key Laboratory of Quantum State Control and Optical Field Manipulation, Department of Physics, Zhejiang Sci-Tech University, 310018 Hangzhou, China}

\date{\today}
\begin{abstract}
Recently, the superconducting properties of Fibonacci quasicrystals have attracted considerable attention. By numerically solving the self-consistent Bogoliubov-de Gennes equations for an $s-$wave superconducting Fibonacci chain, we find that the system exhibits universal end superconductivity, where the pair potential at the chain ends can persist at higher temperatures compared to the bulk critical temperature ($T_{cb}$) of the condensate in the chain center. Furthermore, our study reveals two distinct critical temperatures at the left ($T_{cL}$) and right ($T_{cR}$) ends of the chain. This complex behavior arises from the competition between topological bound states and critical states, a characteristic of quasicrystals. {\color{black}With the chosen parameters, the maximal enhancement of $T_{cR}$ reaches up to $66\%$ relative to $T_{cb}$, while $T_{cL}$ can increase by up to $31\%$.} Our study sheds light on the phenomenon of end superconductivity in Fibonacci quasicrystals, pointing to alternative pathways for increasing the superconducting critical temperature.
\end{abstract}
\maketitle

\section{Introduction}

The Fibonacci chain has attracted great attention since the discovery of three-dimensional icosahedral quasicrystals~\cite{shechtman1984, levine1984} and the realization of GaAs-AlAs heterostructures with alternating layers arranged in a Fibonacci sequence~\cite{merlin1985}. The breakdown of translational invariance in the Fibonacci chain and other quasicrystals leads to multifractal energy spectra with critical states~\cite{suto1989, bellissard1989, grimm2003,mace2016, rai2020, rai2020a, reisner2023, harper1955, aubry1980, lv2022, lv2022a}. These electronic states are distinct from both the extended states in periodic solids and the localized states in strongly disordered solids, being, so to speak, ``in-between"~\cite{rai2020a}, a characteristic highlighted by the term ``critical". Another important property of the Fibonacci chains is the presence of the topologically protected bound states~\cite{huang2018, huang2019b, verbin2013} inherited from the parent crystals (e.g., a 2D square lattice)~\cite{verbin2015}. In general, Fibonacci quasicrystals exhibit a plethora of unusual phenomena. To name a few, in addition to the critical and topological bound states, one can mention the many-body localization~\cite{mace2019}, the hyperuniform states~\cite{torquato2018, baake2019}, reentrant phenomena in the presence of disorder~\cite{jagannathan2020, singh2015}, etc. 

Superconducting properties have also been intensively studied in Fibonacci chains~\cite{rai2019, rai2020, rai2020a, sun2024,wang2024b, kobialka2024a, sandberg2024a}. In particular, for an ensemble of different slices of a fixed length cut from the Fibonacci chain, the disordered gap equation demonstrates that the superconductivity can be enhanced by increasing disorder strength due to the overlap of critical states~\cite{sun2024}. For a one-dimensional tight-binding chain consisting of two parts - a normal section with two Fibonacci-modulated hopping amplitudes and a periodic $s-$wave superconducting section, the self-consistent Bogoliubov-de Gennes (BdG) formalism reveals that topological bound states contribute to superconductivity at every site~\cite{rai2019}. In this case, the pair potential (order parameter) exhibits significant spatial variations with a power-law decay toward the center of the normal section, induced by the multifractal energy spectrum and critical states~\cite{rai2020, rai2020a}. In contrast, for an $s$-wave superconducting Fibonacci chain~\cite{wang2024b} with the Hubbard attractive interaction, the spatial oscillations of the pair potential persist throughout the entire chain and are self-similar. However, the superconducting gap in the local density of states exhibits no spatial variations, being uniform along the chain. Based on these results, it has been claimed~\cite{wang2024b} that as the temperature increases, the superconducting condensate disappears simultaneously at all sites, indicating a unique critical superconducting temperature for the entire Fibonacci chain.

However, recent results obtained for an attractive Hubbard model with the $s$-wave pairing and uniform hopping~\cite{barkman2019, samoilenka2020a, croitoru2020, bai2023, bai2023a, debraganca2023, chen2024} have revealed that the superconducting condensate near the system boundaries (at the chain ends in the one-dimensional case) can persist at temperatures higher than the bulk critical temperature $T_{cb}$. This effect, below referred to as the {\it end superconductivity}, stems from the constructive interference of itinerant quasiparticles, which are not localized near the boundaries but are instead spread throughout the entire system. Moreover, topological bound states can also contribute and further complicate the effect, as revealed for proximitized topological insulators within the Su-Schrieffer-Heeger model~\cite{chen2024c}. Thus, one can expect that the end superconductivity should manifest in the Fibonacci chain - a phenomenon that has been overlooked until now.

In this work, by numerically solving the BdG equations in a self-consistent manner, we investigate the end superconductivity in an $s-$wave superconducting Fibonacci chain. {\color{black}The one-dimensional Fibonacci chain is a reasonable simplification of the three-dimensional Fibonacci quasicrystals, in which periodic two-dimensional lattices are stacked along the third direction in a quasiperiodic way for sufficiently small in-plane hopping amplitudes~\cite{wang2024b}. Notably, these systems are experimentally feasible, as demonstrated by previous experimental results in Refs.~\onlinecite{Todd1986, Cohn1988, Zhu1997}.} Our study reveals that the system exhibits distinct left-end $T_{cL}$ and right-end $T_{cR}$ critical superconducting temperatures, which can far exceed the critical temperature at the chain center $T_{cb}$~{\color{black}(below referred to as the bulk critical temperature for brevity)}. 

The paper is organized as follows. The BdG equations for a Fibonacci chain with $s$-wave superconducting pairing are outlined in Sec.~\ref{II}, along with some details of the numerical algorithm. The results are discussed in Sec.~\ref{III}, which is divided into three subsections. In Section~\ref{III_A}, we investigate the pair potential $\Delta(i)$ together with its single-species quasiparticle contributions. The three critical temperatures ($T_{cR}$, $T_{cL}$ and $T_{cb}$) are considered in Sec.~\ref{III_B}. Finally, Sec.~\ref{III_C} investigates how the three critical temperatures depend on microscopic parameters, such as the chain length and hopping amplitudes. Concluding remarks are given in Sec.~\ref{IV}.

\section{Theoretical Formalism}\label{II}

To numerically investigate the end superconductivity in an $s-$wave superconducting Fibonacci chain, we consider the finite Fibonacci sequence $S_n$, with $n$ being its characteristic sequence  number~\cite{jagannathan2021a}. It contains symbols $A$ and $B$, {\color{black} with the number of symbols} equal to the Fibonacci number $F_n$. For the latter we have $\{F_1,\, F_2,\, F_3,\, F_4,\, F_5, ...\}=\{1,\, 1,\, 2,\,3,\,5, ... \}$. Following the Fibonacci rule, the sequence $S_n$ is the concatenation of $S_{n-1}$ and $S_{n-2}$, i.e., $S_n=[S_{n-1}, S_{n-2}]$, where $S_1=[B]$ and $S_2=[A]$, see Refs.~\onlinecite{zheng1987, jagannathan2021a}. {\color{black} So that, we have $S_3=[AB]$, $S_4=[ABA]$, $S_5=[ABAAB]$, $S_6=[ABAABABA]$ etc.} 

{\color{black}There are two variants of the Fibonacci chain model: the off-diagonal version with the quasiperiodic staggering of the hopping amplitude~\cite{rai2019, wang2024b}, and the diagonal model with the quasiperiodic staggering of on-site potential energy arranged according to the Fibonacci rule, e.g., see Ref.~\onlinecite{sun2024}. These variants have similar results. In our study, we employ the off-diagonal model~\cite{piechon1995, rudinger1998, jagannathan2021a, sun2024} with the hopping amplitudes $t_A$ and $t_B$ arranged in a finite Fibonacci sequence $S_n$. This is achieved by replacing the symbols $A$ and $B$ with $t_A$ and $t_B$, respectively. The number of sites of this chain is $N = F_n + 1$.}

In our study, we utilize the well-known nearest-neighbor tight-binding model, and the superconducting pairing is described~\cite{fu2008,bai2023,bai2023a,wang2024b} by introducing the Hubbard attraction term. The corresponding grand-canonical Hamiltonian reads~\cite{gennes1966, kohmoto1983, ostlund1983, tanaka2000, bai2023a}
\begin{equation}\label{hamilton}
    \mathcal{H}-\mu\mathcal{N}_e = - \sum_{ij}t_{\langle ij\rangle}c_{i\sigma}^\dagger c_{j\sigma} - \sum_{i\sigma}\mu n_{i\sigma}  - g\sum_i n_{i\uparrow}n_{i\downarrow},
\end{equation}
where $\mu$ is the chemical potential, $g >0$ is the on-site attraction coupling, and $\mathcal{N}_e$ is the electron number operator $\mathcal{N}_e= \sum_{i\sigma}n_{i\sigma}=\sum_{i\sigma} c_{i\sigma}^\dagger c_{i\sigma}$, with $c_{i\sigma}^\dagger$ and $c_{i\sigma}$ the creation and annihilation operators of an electron at site $i$ with spin $\sigma=\uparrow,\downarrow$. {\color{black}The summation in Eq.~(\ref{hamilton}) extends over all sites with indexes ($i,j$) from $1$ to $N=F_n+1$. Finally, $t_{\langle ij\rangle}$ stands for the appropriate hopping amplitude between the nearest-neighbor sites: we have either $t_{\langle ij\rangle}=t_A$ or $t_{\langle ij\rangle}=t_B$, following the Fibonacci off-diagonal model.}    

Within the mean-field approximation, we obtain the effective Hamiltonian~\cite{gennes1966}
\begin{equation}
    H_{\rm eff}  = \sum_{ij} {\color{black}H_{ij}} c_{i\sigma}^\dagger c_{j\sigma} 
     + \sum_i[\Delta(i)c_{i\uparrow}^\dagger c_{i\downarrow}^\dagger+ \Delta^*(i)c_{i\downarrow} c_{i\uparrow}],\label{Heff}
\end{equation}
where {\color{black}$H_{ij}=-t_{\langle ij \rangle} - \mu\delta_{ij}$}, $\Delta(i)$ is the superconducting pair potential (the order parameter), and $\delta_{ij}$ is the Kronecker delta function. For simplicity, the Hartree-Fock potential is omitted in Eq.~(\ref{Heff}) since we consider the system at half-filling. In this regime, the spatial electron distribution is uniform (see the results below), and including the Hartree-Fock interaction potential merely shifts the chemical potential. 

Diagonalization of the effective Hamiltonian given by Eq.~(\ref{Heff}), yields the BdG equations~\cite{gennes1966,tanaka2000,samoilenka2020a,bai2023}
\begin{subequations}\label{bdg} 
\begin{align}
 \sum_j {\color{black}H_{ij}} u_\alpha(j) + \Delta(i) v_\alpha(i) &= \varepsilon_\alpha u_\alpha(i), \\
\Delta^*(i) u_\alpha(i) - \sum_j {\color{black}H_{ij}^*} v_\alpha(j) &= \varepsilon_\alpha v_\alpha(i),
\end{align}
\end{subequations}
with $\varepsilon_\alpha$, $u_\alpha(i)$, and $v_\alpha(i)$ being the energy, electron- and hole-like quasiparticle wavefunctions with the quantum number $\alpha$. For convenience, all the positive-energy quasiparticles are arranged in energy ascending order with $\alpha$ the ordering number. The open (quantum confinement) boundary conditions are applied to the quasiparticle wavefunctions: $u_\alpha(i)$ and $v_\alpha(i)$ are set to $0$ at $i=0$ and {\color{black}$i=N+1$}.

The BdG equations given by Eqs.~(\ref{bdg}) are solved in a self-consistent manner, together with the self-consistency condition
\begin{equation}
    \Delta(i)  = g\sum_{\varepsilon_\alpha\ge0} u_\alpha(i)v_\alpha^*(i)(1-2f_\alpha), \label{Delta}
\end{equation}
with $f_\alpha=f(\varepsilon_\alpha)$ being the Fermi-Dirac distribution function. The above sum runs over the physical states with non-zero $\varepsilon_\alpha$. Moreover, the chemical potential $\mu$ in Eqs.~(\ref{bdg}) is determined by the average electron-filling level $\bar{n}_e=\sum_i n_e(i)/N$, with $n_e(i)$ being the spatial electron distribution
\begin{equation}\label{ne}
    n_e(i) = 2\sum_\alpha [f_\alpha|u_\alpha(i)|^2 + (1-f_\alpha)|v_\alpha(i)|^2].
\end{equation} 

Our calculations follow the standard self-consistency numerical routine. First, we construct the BdG equations by introducing initial values of $\mu$ and $\Delta(i)$. Second, $\varepsilon_\alpha$, $u_\alpha(i)$, and $v_\alpha(i)$ are obtained by solving the corresponding eigenvalue problem given by Eqs.~(\ref{bdg}). Then, inserting the solution of this problem into Eq.~(\ref{Delta}), we get a new $\Delta(i)$. In addition, calculating $\bar{n}_e$ and comparing it with its reference number, we increase or decrease $\mu$ by a sufficiently small step. Then, new values of $\Delta(i)$ and $\mu$ are inserted into the BdG equations, and the procedure is repeated until convergence. 

Below, the coupling $g$, the hopping parameter $t_A$, the chemical potential $\mu$, the quasiparticle energy $\varepsilon_\alpha$, and pair potential $\Delta(i)$, are given in units of $t_B$ while the temperature $T$ is in units of $t_B/k_B$, with $k_B$ the Boltzmann constant. In this study we adopt $g=2$, which is a common choice in literature~\cite{tanaka2000, glodzik2018, croitoru2020, bai2023a, chen2024c}. {\color{black}Our iterative procedure continues until both $\Delta(i)$ and $\mu$ converge to within $10^{-10}$. }The chain is considered in the half-filling regime, i.e., $\bar{n}_e$ is set to be $1$. Notably, our qualitative conclusions are not sensitive to the choice of microscopic parameters.

\section{Results and discussions}\label{III}
\subsection{Spatially nonuniform pair potential and end superconductivity}
\label{III_A}

\begin{figure}[t]
\centering
\includegraphics[width=1\linewidth]{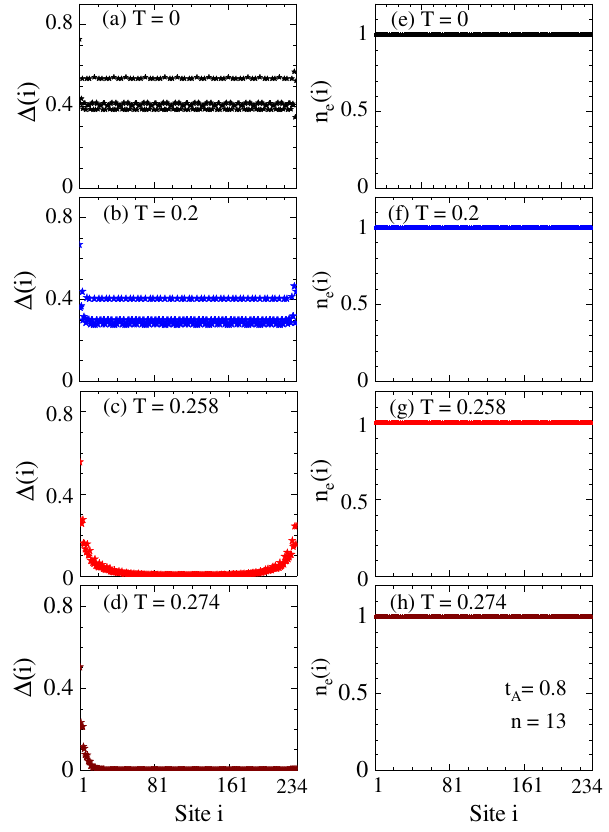}
\caption{The pair potential $\Delta(i)$ and {\color{black}the corresponding spatial electron distribution $n_e(i)$} as functions of the site number $i$ for $T=0$, $0.2$, $0.258$, and $0.274$. The calculations are performed at $t_A=0.8$ for the Fibonacci chain {\color{black}$S_{n=13}$ with the total number of sites $N=F_{13}+1 = 234$}.}
\label{fig1}
\end{figure}

Figures~\ref{fig1}(a-d) illustrate typical spatial profiles of the pair potential $\Delta(i)$ calculated at the temperatures $T=0$, $0.2$, {\color{black}$0.258$, and $0.274$ for the Fibonacci chain $S_{n=13}$} with $t_A=0.8$. As is seen in panels (a) and (b), at $T=0$ and $0.2$, $\Delta(i)$ exhibits remarkable fractal-like behavior inside the chain, which is fundamentally related to the multifractal critical character of the single-electron states in the Fibonacci quasicrystal~\cite{rudinger1998}. {\color{black}At the chain ends $i=1$ and $i=N$, with $N=F_{13}+1=234$, the local pair potential $\Delta(1)$ is remarkably larger than $\Delta(N)$.} These asymmetric condensate distribution is a consequence of the difference in the hopping-amplitude configurations near the left and right chain ends: {\color{black}for $S_{n=13}= [ABAABA\ldots AABAAB]$, we obtain the hopping-amplitude sequence $[t_A\,t_B\,t_A\,t_A\,t_B\,t_A \ldots t_A\,t_A\,t_B\,t_A\,t_A\,t_B]$.}

When increasing $T$ up to {\color{black}$0.258$}, see Fig.~\ref{fig1}(c), the fractal-like pattern of the pair-potential oscillations disappears together with the condensate inside the chain. However, $\Delta(i)$ remains finite near the chain ends so that local maxima of $\Delta(i)$ occur at $i=1$ and {\color{black}$i=N$ with $\Delta(1) =0.555$ and $\Delta(N)=0.243$}. The left-right asymmetry becomes even more pronounced at {\color{black}$T=0.274$}, see Fig.~\ref{fig1}(d). {\color{black}Here $\Delta(N)$ vanishes while $\Delta(1)$ is about $0.503$}: the superconducting condensate survives only at the left end. Thus, we find that the Fibonacci chain exhibits clear signatures of the end superconductivity similar to the interference-induced end superconductivity in the $s-$wave periodic chain~\cite{croitoru2020, bai2023, bai2023a, chen2024}. {\color{black}However, the key difference lies in the spatial symmetry of the condensate distribution: while the periodic case exhibits a superconducting condensate symmetrically distributed about the chain center, the quasiperiodic chain shows an asymmetric distribution.}

{\color{black}The corresponding examples of the spatial electron distribution $n_e(i)$ are shown in Figs.~\ref{fig1}(e-h). One can see that the electron distribution is uniform, and this conclusion holds for any choice of $t_A$ and $n$. Notice that the half-filling case is characterized by the uniform profile $n_e(i)=1$ also in the $s-$wave periodic superconducting chain. Consequently, at half-filling, including the Hartree-Fock term only shifts the chemical potential in both periodic and quasiperiodic systems, leaving the physical results unaffected.}

\begin{figure}[hbt]
\centering
\includegraphics[width=1\linewidth]{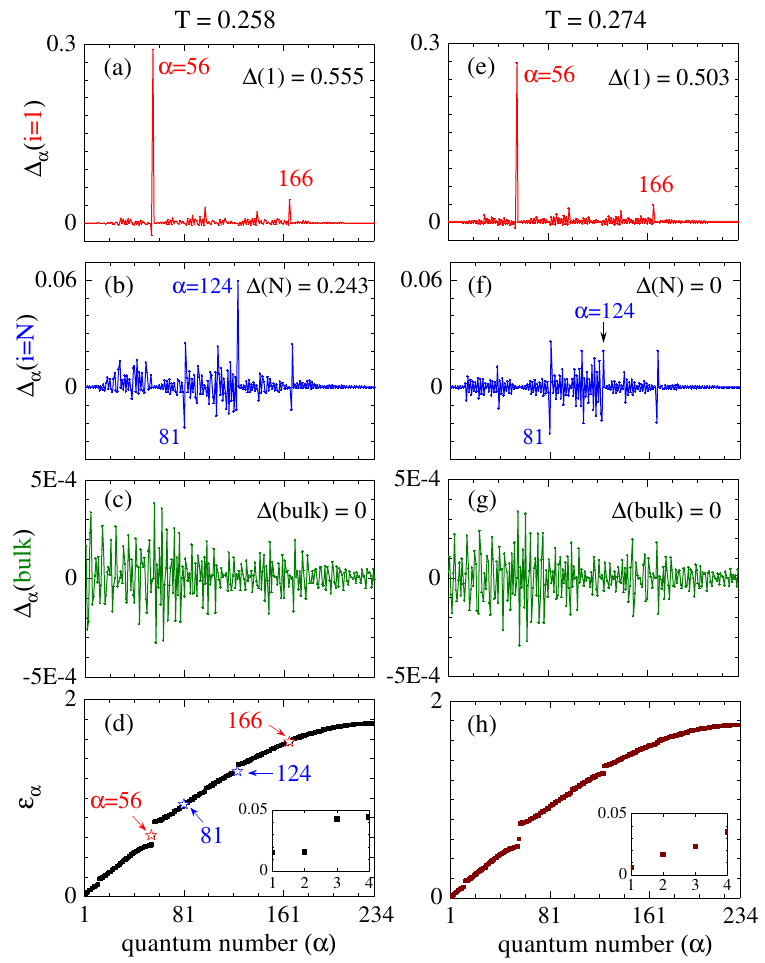}
\caption{(a-c, e-g) Single-species quasiparticle contribution $\Delta_\alpha(i)$ as a function of the quantum number $\alpha$ at $i=1$ (red), $i={\color{black}N}$ (blue) and in the chain center (green) for {\color{black}$T=0.258$ and $0.274$}. (d, h) The quasiparticle energy spectrum $\varepsilon_\alpha$ versus $\alpha$ at {\color{black}$T=0.258$ and $0.274$}, respectively; the insets show a zoomed-in view of the zero-energy region. The calculations are performed for {\color{black}$S_{n=13}$} with $t_A=0.8$.}
\label{fig2}
\end{figure}

To delve deeper, we investigate the single-species contribution of quasiparticles to $\Delta(i)$ given by~\cite{bai2023,bai2023a,chen2024}
\begin{equation}
    \Delta_\alpha(i) = g u_\alpha(i) v^*_\alpha(i)\big(1-2f_\alpha\big).
\end{equation}
Figures~\ref{fig2}(a-b, e-f) show $\Delta_\alpha(i=1,\,N)$ as functions of the quantum number $\alpha$ at {\color{black}$T=0.258$ and $0.274$}, respectively. {\color{black}The calculations are again performed for $S_{n=13}$ with $t_A=0.8$.} For comparison, the single-quasiparticle contribution to the bulk pair potential $\Delta_\alpha({\rm bulk})$ is given versus $\alpha$ in Figs.~\ref{fig2}(c, g) for the same temperatures. The corresponding values of $\Delta(i,\,N)$ and $\Delta({\rm bulk})$ are displayed for the readers' convenience. Notice that $\Delta_\alpha({\rm bulk})$ and $\Delta({\rm bulk})$ are defined as the averages of $\Delta_\alpha(i)$ and $\Delta(i)$ over the range $0.4N \le i \le 0.6N$. The corresponding quasiparticle energies are shown versus $\alpha$ in Fig.~\ref{fig2}(d, h), where the insets represent zoom-ins near zero energy. Only physical quasiparticles (i.e., $\varepsilon_\alpha \geq0$) are considered according to Eq.~({\ref{Delta}}). One can immediately see that there are no zero-energy quasiparticles, which means that the Fibonacci chain with the $s$-wave pairing correlations does not exhibit topological superconductivity.

From Figs.~\ref{fig2}(a, b, e, f), one finds that the most pronounced quasiparticle contributions to {\color{black}$\Delta(i=1, N)$ at $T=0.258$ and $0.274$} depend on the parity of the quantum number $\alpha$: an even (odd) $\alpha$ leads to a positive (negative) contribution. This is related to the fact that both $\Delta_\alpha(i=1)$ and $\Delta_\alpha(i={\color{black}N})$ exhibit rapid oscillations about zero as functions of $\alpha$. Similar zero-crossing oscillations takes place for $\Delta_\alpha ({\rm bulk})$, as seen from Figs.~\ref{fig2}(c,g). Such oscillations appear due to rapid sign variations of $u_\alpha(i)$ and $v_\alpha(i)$ with respect to $\alpha$, which is similar to the previous results obtained for the periodic $s-$wave superconducting chain~\cite{yin2023}. {\color{black}Notably, $\Delta_\alpha (i=1, N)$ are nearly negligible for higher-energy quasiparticles with $\varepsilon_\alpha > 1.57$. However, this is not the case for $\Delta_\alpha({\rm bulk})$.}

\begin{figure*}[htp]
\centering
\includegraphics[width=1\linewidth]{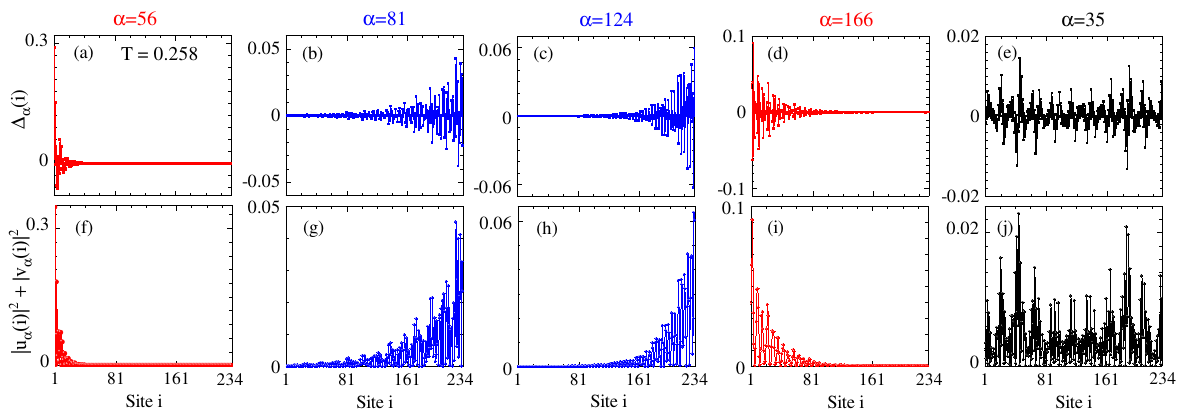}
\caption{(a-e) Typical examples of the spatial profile of the single-species contribution $\Delta_\alpha(i)$ for topological end quasiparticles with {\color{black}$\alpha=56$ and $166$ (in red), $81$ and $124$ (in blue)}, and for the critical quasiparticle state with $\alpha=35$ (in black). (f-j) The corresponding examples of the probability density distribution $|u_\alpha(i)|^2 + |v_\alpha(i)|^2$. {\color{black}The results are calculated for {\color{black}$S_{n=13}$} at $t_A=0.8$ and $T=0.258$}.}
\label{fig3}
\end{figure*}

One can see from Figs.~\ref{fig2}(c, g) that the positive and negative parts of the oscillation pattern of $\Delta_{\alpha}({\rm bulk})$ are nearly mirror images of each other. This behavior stems from the fact that positive and negative contributions to $\Delta({\rm bulk})$ are the same for both highlighted temperatures, as $\Delta({\rm bulk})=0$. Remarkably, while the individual quasiparticle contributions are non-zero in this case, their sum vanishes - a behavior that shows clear similarities to destructive interference phenomena. {\color{black}In addition, it is seen from Fig.~\ref{fig2}(f) that $\Delta_{\alpha}(N)$ exhibits a similar $\alpha$-dependence at $T=0.274$, as $\Delta(i=N) = 0$ at $T=0.274$.}

{\color{black}The behavior of $\Delta_{\alpha}(i=1)$ and $\Delta_{\alpha}(i=N)$ in Figs.~\ref{fig2}(a, b, e) differs qualitatively from the pattern discussed in the previous paragraph. In this case, the positive individual quasiparticle contributions predominate in the oscillations so that their sum is nonzero, which is similar to constructive interference phenomena. As a results, we have $\Delta(1)=0.555$ and $\Delta(N)=0.243$ at $T=0.258$, and $\Delta(1)=0.503$ at $T=0.274$. More specifically, at $T=0.258$, two most pronounced contributions to the left-end pair potential are $\Delta_{56}(1)=0.52\Delta(1)$ and $\Delta_{166}(1)=0.07\Delta(1)$. At the left end, the most significant single-species contribution to the order parameter is positive and given by $\Delta_{124}(N)=0.25\Delta(N)$. However, there are also remarkable negative single-species contributions, for example, $\Delta_{81}(N)=-0.09\Delta(N)$, which is a reflection of the fact that $\Delta(N)$ is notably smaller than $\Delta(i=1)$ at $T=0.258$. At $T=0.274$ only the left-end order parameter remains nonzero, as seen from Fig.~\ref{fig2}(e). Here, similarly to the case of $T=0.258$, there are two most important contributions to the pair potential given by $\Delta_{56}(1)=0.53\Delta(1)$ and $\Delta_{166}(1)=0.06\Delta(1)$.}

The quasiparticle energies for $\alpha={\color{black}56}$, $81$, $124$ and {\color{black}$166$} are highlighted in Fig.~\ref{fig2}(d). One can see that the energy levels for $\alpha={\color{black}56}$ and $124$ are located at the band edges, near large energy gaps. The single-electron states associated with these quasiparticles represent topological bound states of the normal Fibonacci chain, with energies lying between neighboring bands~\cite{jagannathan2021a}. In this work, the corresponding quasiparticles are referred to as {\it topological bound quasiparticles} or, more specifically, {\it topological end states} when describing quasiparticles localized near the chain ends. These states are similar to the quasiparticles corresponding to the topological bound states in a proximitized Su-Schrieffer-Heeger chain~\cite{chen2024}. The states with $\alpha = 81$ and {\color{black}$166$} are also topological bound quasiparticles, but their energies are located at the band edges near almost insignificant band gaps not visible in the figure. 

{\color{black}For further details, Figs.~\ref{fig3}(a-d, f-i) demonstrate spatial profiles of $\Delta_\alpha(i)$ and the probability density distribution $|u_\alpha(i)|^2 +|v_\alpha(i)|^2$ for the topological end quasiparticles with $\alpha={\color{black}56}$ (a, f), $81$ (b, g), $124$ (c, h), and ${\color{black}166}$ (d, i) at {\color{black}$T=0.258$}. The calculations are done for the same Fibonacci chain $S_{n={\color{black}13}}$ with $t_A=0.8$. In panels (a) and (d), both $\Delta_{{\color{black}56}}(i)$ and $\Delta_{{\color{black}166}}(i)$ exhibit pronounced rapid oscillations near the left end of the chain while vanishing at its right end. $\Delta_{{\color{black}56}}(i)$ decays toward the chain center with a much shorter characteristic length than $\Delta_{{\color{black}166}}(i)$. Similarly, the corresponding quasiparticle probability density distributions $|u_\alpha(i)|^2 +|v_\alpha(i)|^2$, given in Fig.~\ref{fig3}(f), are pronounced near the left end while decaying toward the chain center and vanishing at the right end. Notice that the characteristic spatial decay length of $|u_\alpha(i)|^2 +|v_\alpha(i)|^2$ is similar to that of $\Delta_{\alpha}(i)$. }

{\color{black}In turn, for the topological bound states with $\alpha=81$ and $124$, both $\Delta_\alpha(i)$ in Figs.~\ref{fig3}(b, c) and $|u_\alpha(i)|^2 + |v_\alpha(i)|^2$ in Figs.~\ref{fig3}(g, h) exhibit significant oscillations near the right end of the chain while decaying toward the left end. Both states are specified by similar spatial lengths, which are much larger than the characteristic length of the state with $\alpha=56$ but comparable with that of $\alpha=166$. }

In Figs.~\ref{fig3}(e, j), $\Delta_{\alpha}(i)$ and $|u_\alpha(i)|^2+|v_\alpha(i)|^2$ are given versus the site number $i$ for the quasiparticle state with $\alpha=35$, representing the critical single-electron states responsible for the fractal-like behavior of the pair potential inside the chain, see, e.g., Refs.~\cite{mace2016, rai2020, rai2020a, reisner2023}. Below, this kind of quasiparticles is referred to as {\it critical quasiparticles}. One can see that, contrary to the results for the topological bound quasiparticles, the oscillations in $\Delta_{35}(i)$ and $|u_{35}(i)|^2+|v_{35}(i)|^2$ are spread throughout the entire chain. {\color{black}For example, we obtain $\Delta_{35}(1)=-0.0016$ and $\Delta_{35}(N)=0.0065$.} In addition, the probability density distribution of these quasiparticles has a clear multifractal character~\cite{kohmoto1987}. 

\begin{figure}[t]
\centering
\includegraphics[width=1\linewidth]{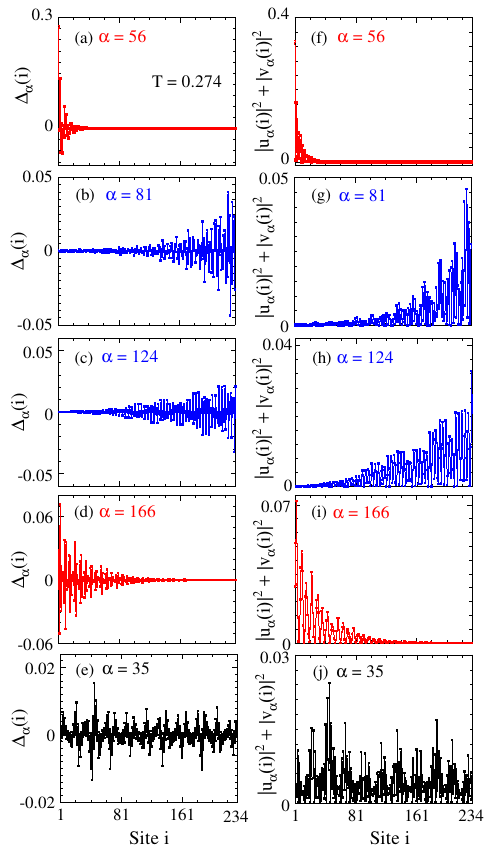}
\caption{The same as in Fig.~\ref{fig3} but at {\color{black}$T=0.274$}.}
\label{fig4}
\end{figure}

Now, let us consider {\color{black}$T= 0.274$}. The corresponding examples of $\Delta_\alpha(i)$ and $|u_\alpha(i)|^2 +|v_\alpha(i)|^2$ are shown as functions $i$ in Fig.~\ref{fig4} for the topological bound quasiparticles with {\color{black}$\alpha=56$~(a, f), $81$~(b, g), $124$~(c, h), $166$~(d, i)}, and for the critical quasiparticle state with {\color{black}$\alpha=35$}~(e, j). As previously, we investigate $S_{n={\color{black}13}}$ with $t_A=0.8$. {\color{black}One finds that the results for $\Delta_\alpha(i)$ and $|u_\alpha(i)|^2 +|v_\alpha(i)|^2$ at $\alpha=56$ and $81$ are nearly the same as the corresponding calculations for $T= 0.258$. The data for $\alpha=126$ and $166$ are similar to their counterparts for $T=0.258$, but the spatial localization lengths for these states are larger at $T=0.274$. }

{\color{black}The critical quasiparticle state with $\alpha=35$ is spread throughout the whole chain at $T=0.274$, similar to the spatial distribution of this state at $T=0.258$. Its contributions to the left-end and right-end pair potentials are not negligible. For example, we find $\Delta_{35}(1)=-0.0020$ and $\Delta_{35}(N)=0.0034$ at $T=0.274$. In general, the total contribution of the critical states to the end-superconductivity is comparable to that of the topological bound states. }

{\color{black} Here we note that the left-end and right-end condensates remain connected even when the central region has zero condensate density. This connection is mediated by critical states that extend across the entire chain, contributing to both ends simultaneously. These delocalized states prevent solutions with opposite-sign pair potentials at the opposite ends.}

Thus, similar to the previous results~\cite{croitoru2020, chen2022, bai2023, bai2023a, debraganca2023} obtained for a periodic $s-$wave superconducting chain, we find the formation of an inhomogeneous condensate distribution with the pronounced end enhancements in the Fibonacci chain $S_{n={\color{black}13}}$ due to the interference-like features manifesting in the summation of individual quasiparticle contributions to the pair potential. However, notably, now the phenomenon is accompanied by a complex interplay between topological bound quasiparticles and non-localized critical quasiparticle states. This bears some resemblance to the recent results for proximitized topological insulators described by the Su-Schrieffer-Heeger model~\cite{chen2024}. 

\subsection{Three critical temperatures}\label{III_B}

Now, {\color{black}we consider the problem of multiple superconducting critical temperatures in the Fibonacci chain}. As is seen in Figs.~\ref{fig1}(c, d), the superconductivity in the chain $S_{n={\color{black}13}}$ with $t_A=0.8$ is essentially inhomogeneous, i.e., the end and bulk pair potentials $\Delta(i=1,\,{\color{black}N},\,{\rm bulk})$ drop to zero at different temperatures. As a result, one obtains distinct critical temperatures for the end and bulk regions, similar to the conclusions reached for an $s$-wave periodic superconducting chain~\onlinecite{croitoru2020, chen2022,bai2023, bai2023a,debraganca2023}. Moreover, one can see that the pair potential at the left end of the Fibonacci chain in Fig.~\ref{fig1} persists at higher temperatures compared to the right end. In this way, we observe two distinct end critical temperatures. Notice that such a feature does not exist in the periodic case, see Refs.~\onlinecite{croitoru2020, chen2022,bai2023, bai2023a,debraganca2023}. In this section, we investigate how the bulk $T_{cb}$, the left-end $T_{cL}$, and right-end $T_{cR}$ critical temperatures are sensitive to the sequence number $n$ and hopping coefficient $t_A$~(in units of $t_B$). 

\begin{figure}[t]
\centering
\includegraphics[width=1\linewidth]{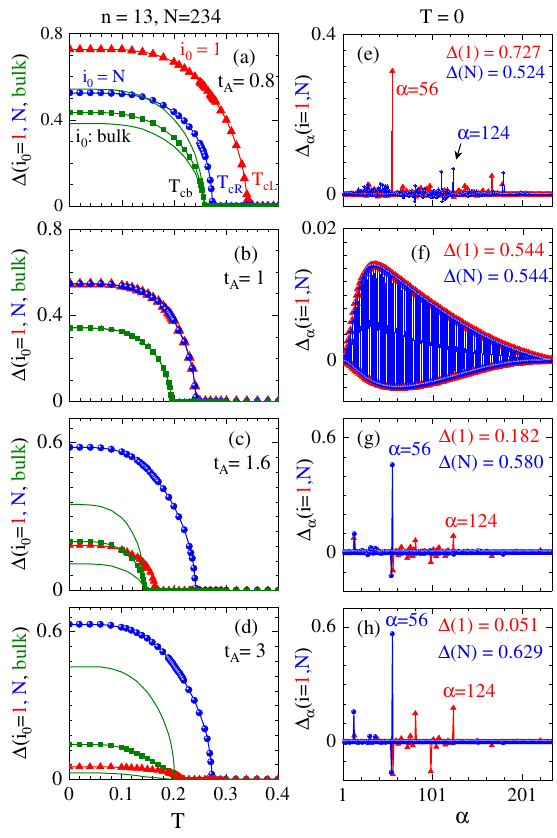}
\caption{(a-d) Temperature-dependent pair potentials $\Delta(1)$~(blue circles), $\Delta(N)$~(red triangles) and $\Delta({\rm bulk})$~(green squares) for the chain {\color{black}$S_{n=13}$ with $t_A=0.8$, $1$, $1.6$, and $3$}. {\color{black}In each panel, the two solid green curves represent the maximal and minimal values of the multifractal order parameter, calculated over the site domain $[0.4N,\,0.6N]$.} The critical temperatures $T_{cL}$, $T_{cR}$, and $T_{cb}$ are defined in panel (a): each corresponds to the temperature at which the respective pair potential drops to zero. (e-h) $\Delta_{\alpha}(i=1,\,{\color{black}N})$ versus $\alpha$ at $T=0$ for the same chain and $t_A$ values. {\color{black}For reference, the end pair potentials $\Delta(1)$ and $\Delta(N)$ are also included.}}
\label{fig5}
\end{figure}

The $T$-dependent pair potential $\Delta(i)$ calculated at $i=1$, $N$, and in the chain center (averaged over the range $0.4N \le i \le 0.6N$) for $S_{n={\color{black}13}}$ is shown in Figs.~\ref{fig5}(a-d) for $t_A=0.8$, $1$, {\color{black}$1.6$, and $3$}, see the curves marked by triangles (red), spheres (blue), and squares (green), respectively. {\color{black}The two solid green curves shown in Figs.~\ref{fig5}(a-d) exhibit the maximal and minimal values of the multifractal pair potential in the central region $0.4N \le i \le 0.6N$. To better understand the microscopic picture, Figs.~\ref{fig5}(e-h) illustrate the corresponding single-species quasiparticle contributions $\Delta_\alpha(i=1,N)$ as functions of $\alpha$ at $T=0$; the data for $i=1$ are marked by red triangles while the results for $i=N$ are given by blue circles. In general, one can see that the $T-$dependent data for $\Delta(i)$ are qualitatively similar to those of the conventional BCS theory~\cite{gennes1966}. However, numerically, they deviate from the BCS trend.}

From Figs.~\ref{fig5}(a-d) one finds that $\Delta(i=1,\,N,\,{\rm bulk})$ and the critical temperatures $T_{cR}$, $T_{cL}$, and $T_{cb}$ are very sensitive to $t_A$. In particular, for the case of $t_A=0.8$ illustrated in Fig.~\ref{fig5}(a), we find $T_{cL}>T_{cR}>T_{cb}$, {\color{black}with $T_{cL}=0.346$, $T_{cR}=0.274$ and $T_{cb}=0.258$.} This end superconductivity with the breakdown of the left-right symmetry is determined by the competition between the topological bound and critical quasiparticles, as discussed in the previous subsection. 

\begin{figure}[t]
\centering
\includegraphics[width=1\linewidth]{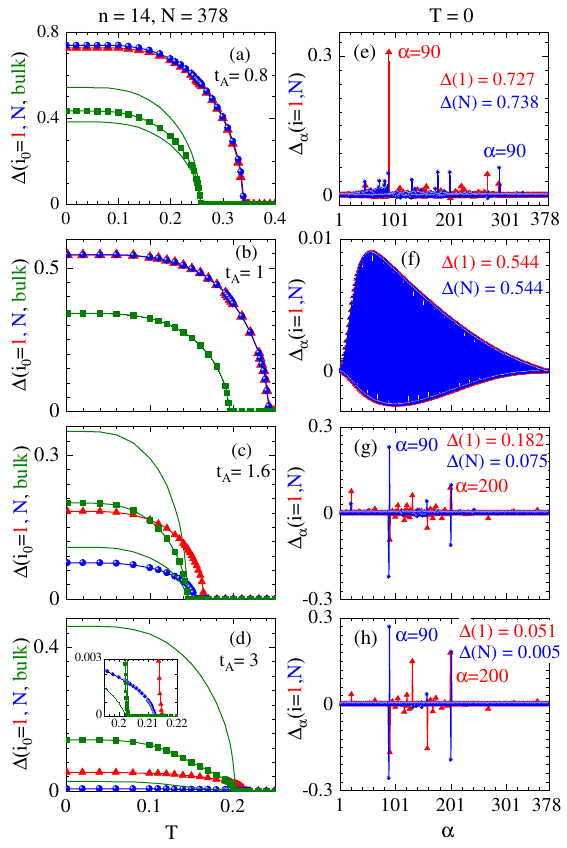}
\caption{The same as in Fig.~\ref{fig5} but for the Fibonacci chain {\color{black}$S_{n=14}$ with $N=378$}. {\color{black}The inset in panel (d) shows the zoomed-in view of the left-end, right-end, and bulk pair potentials in the temperature range $[0.195,\,0.22]$, including $T_{cL}$, $T_{cR}$, and $T_{cb}$.}}
\label{fig6}
\end{figure}

For $t_A=1$ in Fig.~\ref{fig5}(b), the system is reduced to a periodic $s-$wave superconducting chain without critical and topological end states~\cite{bai2023}. As a result, the left- and right-end pair potentials $\Delta(1)$ and $\Delta(N)$ become the same, and we arrive at the results obtained previously in Refs.~\onlinecite{bai2023, croitoru2020}. {\color{black}In this case, one finds $T_{cL}=T_{cR} > T_{cb}$. It is of importance to mention that the $\alpha$-dependence of $\Delta_\alpha(i=1,\,N)$ for $t_A=1$ is qualitatively different from that for $t_A=0.8$, cf. Figs.~\ref{fig5}(e) and Fig.~\ref{fig5}(f). The dramatic contrast between Fig. 5(f) and Figs. 5(e) stems from topological bound states, which generate prominent peaks in $\Delta_{\alpha}(1)$ and $\Delta_{\alpha}(N)$ as functions of $\alpha$ for $t_A \not = 1$. If the contributions of topological bound states were removed from Figs.~5(g,e), the resulting pattern would resemble Fig.~5(f), albeit with some modifications due to critical states. However, this similarity is not immediately obvious because $\Delta_{\alpha}(1)$ and $\Delta_{\alpha}(N)$ are confined to the range $[0, 0.02]$ in Fig. 5(f), while in Figs. 5(g,e) they span $[0,0.4]$. As $t_A \to t_B$, the contributions of topological bound states gradually vanish, and the system transitions to the periodic case without abrupt changes in the $\Delta_{\alpha}(1,\,N)$ distribution.}

{\color{black}For the case of $t_A=1.6$, illustrated in Fig.~\ref{fig5}(c), $\Delta(1)$ is smaller than $\Delta(N)$ at any temperature. As a consequence, we obtain $T_{cL} < T_{cR}$. At sufficiently low temperatures $\Delta(1)$ is also smaller than the averaged bulk pair potential $\Delta({\rm bulk})$, suggesting that $T_{cL}$ should be smaller than $T_{cb}$. However, this is not the case as $\Delta(1)$ exceeds $\Delta({\rm bulk})$ at higher temperatures. For $t_A=1.6$ the topological quasiparticle contribution with $\alpha=56$ predominates in the right-end pair potential. For example, at $T=0$ we find $\Delta_{56}(N)=0.79\Delta(N)$, as shown in Fig.~\ref{fig5}(g). The left-end condensate is also mainly determined by topological bound states - for instance, the quasiparticle state with $\alpha=124$ contributes significantly to the left-end pair potential. }

{\color{black}When $t_A$ increases to $3$, see Fig.~\ref{fig5}(d), we obtain $T_{cR} = 0.276$, $T_{cL}=0.216$, and $T_{cb}=0.204$. Similar to the case $t_A=1.6$, $\Delta(1)$ is smaller than $\Delta({\rm bulk})$ at low temperatures, yet $T_{cL} > T_{cb}$. The right-end pair potential is dominated by topological bound states. At $T=0$, the primary contribution to the right-end pair potential comes from the topological bound quasiparticle state with $\alpha=56$, where $\Delta_{56}(N)=0.9\Delta(N)$. For the left-end pair potential, there are four topological quasiparticle states ($\alpha=57$, $82$, $99$, and $124$) with significant individual contributions. However, their contributions have alternating signs: at $T=0$, we obtain $\Delta_{57}(1)=-0.169$, $\Delta_{82}(1)=0.148$, $\Delta_{99}(1)=-0.154$ and $\Delta_{124}(1)=0.179$. The net sum of these contributions ($0.004$) is much smaller than the total left-end pair potential $\Delta(1)=0.051$, indicating that critical quasiparticles play a significant role here.}

Now we consider {\color{black}the Fibonacci chain $S_{n=14}$ with $N=378$}. For illustration, we plot $\Delta(i=1,\,{\color{black}N},\,{\rm bulk})$ as functions of $T$ in Figs.~\ref{fig6}(a-d) for $t_A = 0.8$, $1$, {\color{black}$1.6$ and $3$}, respectively. In Figs.~\ref{fig6}(e-h), one can find $\Delta_\alpha(i=1,\,N)$ as functions of $\alpha$ at $T=0$ for the same set of $t_A$ values. {\color{black}The results for $\Delta(i=1,\,{\rm bulk})$ in Figs.~\ref{fig6}(a-d) match those in Figs.~\ref{fig5}(a-d), indicating that  $T_{cL}$ and $T_{cb}$ are identical for $n=13$ and $n=14$~(for corresponding  $t_A$ values, of course). Based on this observation, we can expect that $\Delta(i=1,\,{\rm bulk})$, $T_{cL}$, and $T_{cb}$ are independent of $n$. This $n$-invariance arises because the left-end chain configuration remains unchanged for sufficiently large $n$. Similarly, the chain arrangement in the domain $[0.4N, 0.6N]$, used to calculate $\Delta({\rm bulk})$, repeats under the Fibonacci chain construction routine for sufficiently long chains. As an important remark, one should not be misled by the differing quantum numbers of the quasiparticles that contribute most significantly to $\Delta(1)$ for $n=13$ and $n=14$, cf. Figs.~\ref{fig5}(h) and \ref{fig6}(h).} In our numerical procedure, the quasiparticle states are enumerated in the ascending energy order, meaning that states sharing the same index $\alpha$ represent different physical states for different $n$.

When further comparing Figs.~\ref{fig5} and \ref{fig6}, one can see that, {\color{black}unlike the left-end and bulk chain characteristics, $\Delta(N)$ and $T_{cR}$ are particularly sensitive to the choice of $n$. This behavior stems from differences in the right-end configurations between $S_{n=13}$ and $S_{n=14}$. Specifically, the last three hopping amplitudes at the right end are $t_At_At_B$ for $n=13$ but $t_At_Bt_A$ for $n=14$. These structural differences lead to significant variations in the right-end pair potential between $n=13$ and $n=14$, cf. Figs.~\ref{fig5} and \ref{fig6}. The most remarkable difference occurs at $t_A=3$, where $T_{cR}=1.28T_{cL}$ for $n=13$, see Fig.~\ref{fig5}(d), while $T_{cR} \approx T_{cL}$ for $n=14$, see Fig.~\ref{fig6}(d). }

One can also see {\color{black}in Fig.~\ref{fig5}(h) that for $n=13$ and $t_A=3$, the most significant input to $\Delta(N)$ at $T=0$ comes from the quasiparticle state with $\alpha=56$. In contrast, Fig.~\ref{fig6}(h) reveals that for $n=14$, four quasiparticle states contribute significantly to $\Delta(N)$ at $T=0$: these are quasiparticles with $\alpha=89,\, 90,\, 201$, and $202$. Notably, these contributions alternate in sign, leading to a sharp reduction in $T_{cR}$, as compared to the case of $n=13$.}

\subsection{Dependence of the critical temperatures on the hopping parameter $t_A$ and the characteristic sequence number $n$}
\label{III_C}

\begin{figure}[t]
\centering
\includegraphics[width=1\linewidth]{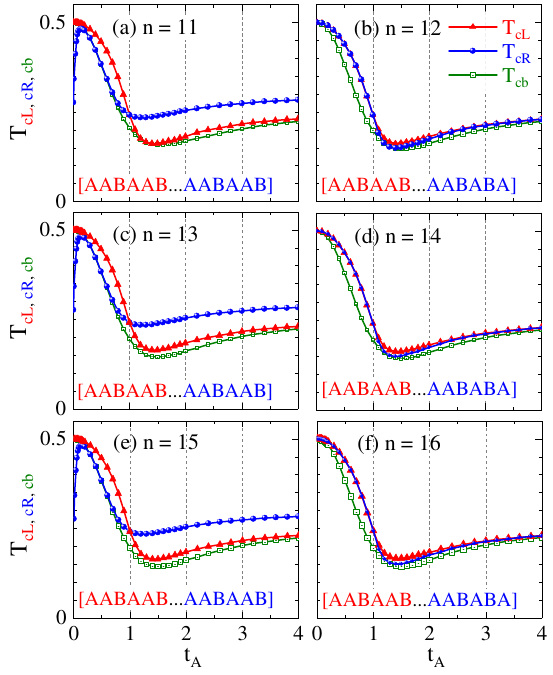}
\caption{The critical temperatures $T_{cL}$ (blue spheres), $T_{cR}$ (red triangles) and $T_{cb}$ (green squares) as functions of $t_A$ for the chains $S_{n}$ with $n=11$~(a), $12$~(b), $13$~(c), $14$~(d), $15$~(e), and $16$~(f). The left (right) panels represent odd (even) $n$. In each panel, six symbols at the chain end--controlling the hopping amplitudes--are highlighted. The chain lengths for $n=11,\,12, ...,\,16$ are $N=90$, $145$, $234$, $378$, $611$, and $988$, respectively.}
\label{fig7}
\end{figure}

Here, we conduct a systematic study of how the parameters $t_A$ and $n$ influence the end and bulk critical temperatures, i.e., $T_{cR}$, $T_{cL}$, and $T_{cb}$. Figure~\ref{fig7} displays these critical temperatures as functions of $t_A$ for {\color{black}$n=11,\,...,\,16$}, with the left (right) panels corresponding to odd (even) $n$. $T_{cL}$ and $T_{cR}$ are represented by red triangles and blue spheres, respectively, while $T_{cb}$ is shown by green squares. In all panels, the chain configurations on the left (red) and right (blue) ends of $S_n$ are highlighted. 

{\color{black}Consistent with the discussion of Sec.~\ref{III_B}, Fig.~\ref{fig7} reveals that $T_{cb}$~(green squares) remains independent of $n$ -- except for $n=11$, where slight deviations occur due to finite-size effects, particularly near $t_A=1$. The bulk critical temperature reaches its minimum $T_{cb}=0.162$ at $t_A=1.5$, which is close to $t_A=1$ corresponding to the ordinary attractive Hubbard model for the periodic chain.} As {\color{black}$t_A$ deviates from $1.5$}, $T_{cb}$ grows significantly in agreement with the previous results of the disordered gap equation~\cite{sun2024}. Notice that $T_{cb}$ is also dependent on other parameters, e.g., the chemical potential $\mu$ and the coupling strength $g$~\cite{wang2024b}. However, investigations of these aspects are beyond the scope of our work. 

{\color{black}One can also see that $T_{cL}$~(red triangles) is independent of the sequence number $n$ in Fig.~\ref{fig7}, which agrees with the results of Figs.~\ref{fig5} and \ref{fig6}. As discussed in the last paragraph of Sec.~\ref{III_B}, this occurs because the left-end configuration remains the same for chains with different $n$. The first six symbols at the left end for $n=11,\,...,\,16$ are the same and given by the sequence $ABAABA$ displayed in Figs.~\ref{fig7}(a-f). We recall that the corresponding sequence of the hopping amplitudes is obtained by replacing symbols $A$ and $B$ by $t_A$ and $t_B$, respectively. }

{\color{black}As can be expected from the results discussed in Sec.~\ref{III_B}, the right-end critical temperature $T_{cR}$~(blue spheres) exhibits significant variations with $n$. This expectation is confirmed by the data given in Figs.~\ref{fig7}(a-f). However, more precisely, we find that the behavior of $T_{cR}$ versus $t_A$ depends on the parity of $n$. This is certainly related to the fact that the first six symbols on the right end of the chain are determined by this parity: $AABAAB$ for odd $n$, see Figs.~\ref{fig7}(a, c, e); $AABABA$ for even $n$, see Figs.~\ref{fig7}(b, d, f).} 

{\color{black}There are two important parity-dependent features of our results shown in Fig.~\ref{fig7}. First, for $t_A \to 0$, $T_{cR}$ is significantly smaller than $T_{cL}$ for odd $n$, while $T_{cR}$ and $T_{cL}$ are the same for even $n$. This behavior stems from the system’s fragmentation at $t_A$, where the Fibonacci chain (for any $n$) breaks into isolated short segments due to vanishing hopping at their boundaries. These segments consist of either single sites or two-site clusters connected by $t_B$. The quantum confinement impact is more pronounced in single isolated sites than in two-site segments. Consequently, the pair state on a single site is more robust compared to that on a two-site cluster. For odd $n$, the sequence $AABAAB$ appears as both the first six symbols at the left end and the last six symbols at the right end of the chain, as shown in Figs.~\ref{fig7}(a, c, e). Therefore, when $t_A=0$, the left-end superconductivity is governed by a single isolated site while the right-end superconductivity depends on a two-site cluster. Following our earlier discussion of quantum confinement effects, this arrangement naturally leads to $T_{cL} > T_{cR}$, consistent with the results in Figs.~\ref{fig7}(a, c, e). 

For even $n$, see Figs.~\ref{fig7}(b, d, f), the right-end sequence changes to $AABABA$ while the left-end configuration remains identical to the odd $n$ case. In this scenario, at $t_A=0$, both ends feature isolated single-site segments controlling the corresponding pair potentials, which results in $T_{cR}=T_{cL}$. This equality matches perfectly with the data presented in Figs.~\ref{fig7}(b, d, f). 

The second key distinction between the results for odd and even $n$ emerges for $t_A \geq 2$. Here, for odd $n$, $T_{cR}$ exceeds $T_{cL}$ while for even $n$, $T_{cR}$ nearly equals to $T_{cL}$. As established earlier from Figs.~\ref{fig5}(d) and \ref{fig6}(d), the enhancement of $T_{cR}$ at $t_A\geq2$ for odd $n$ originates from the predominating contribution of a single species of topological bound states to the right-end pair potential, see Fig.~\ref{fig5}(h). Conversely, for even $n$, multiple quasiparticle bound states exist [see Fig.~\ref{fig6}(h)], with their contributions to $\Delta(i=N)$ alternating in sign, resulting in a suppressed $T_{cR}$ relative to $T_{cL}$.  

It is important to note that while $T_{cL}$ in Fig.~\ref{fig7} is generally higher than $T_{cb}$~(except specific points where $T_{cL}\approx T_{cb}$), $T_{cR}$ can be smaller than $T_{cb}$. In particular, we find that $T_{cR} < T_{cb}$ for odd $n$ when $t_A <0.2$. Nevertheless, the most significant enhancement of end superconductivity in Fibonacci chains appears at the right end for odd $n$ and $t_A \gtrsim 1$. }

\begin{figure}[t]
\centering
\includegraphics[width=1\linewidth]{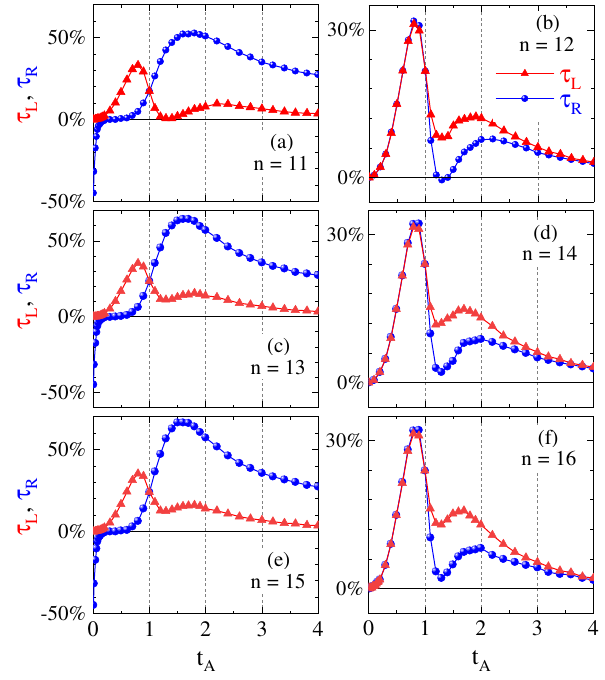}
\caption{The relative enhancements $\tau_{R,L} = (T_{cR,cL}-T_{cb})/T_{cb}$ as functions of $t_A$ for the Fibonacci chains $S_{n}$ with {\color{black}$n=11,\,...,\,16$}. In each panel, the black horizontal solid line represents $\tau_{R,L}=0$.}
\label{fig8}
\end{figure}

Additional details can be found in Fig.~\ref{fig8}, which plots the relative quantities $\tau_{L} = (T_{cL}-T_{cb})/T_{cb}$ and $\tau_{R} = (T_{cR}-T_{cb})/T_{cb}$ versus $t_A$ {\color{black}for $n=11,\, ...,\,16$, demonstrating how the end critical temperatures can be enhanced relative to the bulk superconducting temperature. Here, red triangles represent $\tau_L$, whereas blue spheres denote $\tau_R$. The left panels (a, c, e) in Fig.~\ref{fig8} display results for odd $n$ while the right panels (b, d, f) correspond to even $n$. Here we see that the finite-size effects, while clearly present for $n=11$ and $12$, are washed out for larger $n$. Ignoring the finite-size deviations, we find that for both parities, the left-end superconducting temperature exceeds the bulk critical temperature in the range $t_A > 0.2$, with the maximum $\tau_L=33\%$ occurring at $t_A\approx0.9$. The right-end enhancement occurs under different conditions depending on the parity: for odd $n$ the enhancement appears when  $t_A > 0.8$; for even $n$ the enhancement occurs for all  $t_A > 0$. In turn, the maximum enhancement values also show the clear parity dependence: for odd $n$, $\tau_R$ reaches $65$-$68\%$ at $t_A\approx1.6$; for even $n$, $\tau_R$ attains a maximum of about $33\%$ at $t_A\approx0.9$.}

\begin{figure}[t]
\centering
\includegraphics[width=0.8\linewidth]{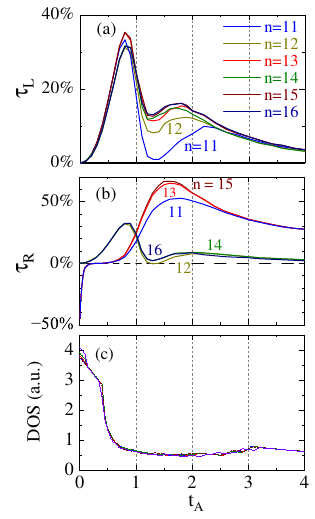}
\caption{For the Fibonacci chains $S_n$ with $n = 11, ..., 16$: (a,b) $\tau_L$ and $\tau_R$ as functions of $t_A$; the dashed horizontal line in panel (b) displays $\tau_R=0$. (c) Density of states (DOS) at the chemical potential $\mu$ as a function of $t_A$. The DOS is calculated as $\frac{{\rm \Delta}N_s}{N{\rm \Delta}E}|_{E=\mu}$ at $g=0$, with ${\rm \Delta}N_s$ the number of states in the energy interval $\mu-\hbar\omega_D\le E\le\mu+\hbar\omega_D$ and $\delta E=2\hbar\omega_D$. The chain lengths for $n=11, ...,\,16$ are $N=90$, $145$, $234$, $378$, $611$, and $988$, respectively.}
\label{fig9}
\end{figure}
{\color{black} To illustrate the influence of quantum-size effects on surface superconductivity, Fig.~\ref{fig9} presents $\tau_L$ (a) and $\tau_R$ (b) as functions of $t_A$ for $S_n$ chains of lengths $n = 11$ to $16$. Figure~\ref{fig9}(a) shows that $\tau_L$ exhibits nearly identical behavior for $n \geq 13$. In contrast, shorter chains ($n=11, 12$) show clear deviations from this behavior within the range $1 \le t_A \le 2.5$.
For $\tau_R$, shown in Fig.~\ref{fig9}(b), the results are highly sensitive to the parity of $n$. For odd $n$, these results are nearly independent of chain length for $n \geq 13$, while for even $n$, $\tau_R$ exhibits nearly the same behavior for $n \geq 12$. These observations indicate that the features of surface superconductivity stabilize for chains with $n \geq 13$, suggesting that the results for $\tau_L$ and $\tau_R$ converge beyond this length.}

Now, we study the impact of the density of states (DOS) on $T_{cb}$ in Fibonacci chains. Figure~\ref{fig9}(c) plots the DOS, defined as {\color{black}$\frac{{\rm \Delta}N_s}{N{\rm \Delta}E}|_{E=\mu}$, with ${\rm \Delta}N_s$ being the number of states in the energy interval $\mu-\hbar\omega_D\le E\le\mu+\hbar\omega_D$ and $\Delta E=2\hbar\omega_D$}. It is given as a function of $t_A$ and calculated for $n=11,\,...,\,16$ at $g=0$. As seen from {\color{black}Fig.~\ref{fig9}(c)}, increasing $t_A$ initially causes the DOS to drop to its minimum value of approximately $0.5$ at $t_A\approx1.4$. The DOS then remains nearly constant across all $n$ for $1.2<t_A<2.6$, followed by a slight increase in the range $2.7<t_A<3.05$. For $t_A>3.05$, the DOS decreases almost linearly. Comparing these results with the results for $T_{cb}$~(green squares) in Fig.~\ref{fig7}, one concludes that $T_{cb}$ follows the DOS in the region $0<t_A<1.4$: the larger the DOS, the higher the critical temperature. This is qualitatively consistent with the textbook BCS scenario~\cite{gennes1966, ketterson1999}. However, at $t_A>3.05$, $T_{cb}$ and the DOS exhibit opposite trends: $T_{cb}$ saturates while the DOS decreases. Our analysis indicates that this behavior arises from an enhanced contribution of topological bound states - a mechanism independent of the DOS, in contrast to the standard BCS framework (and unlike the contribution from critical states). The increasing influence of topological bound quasiparticles compensates for the diminishing DOS effect, leading to the observed saturation of $T_{cb}$. We propose that similar physics governs the saturation of both $T_{cL}$ and $T_{cR}$ for $t_A > 3$.

\begin{figure}[t]
\centering
\includegraphics[width=0.8\linewidth]{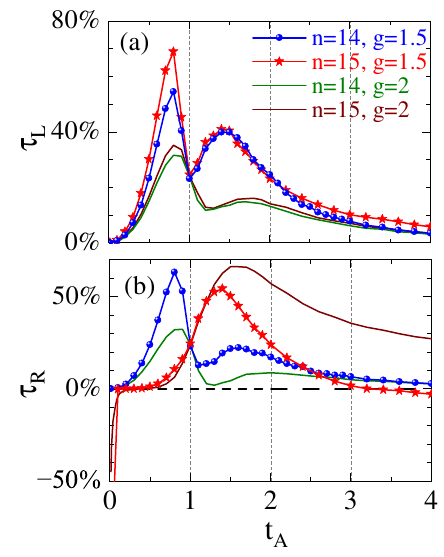}
\caption{\color{black}Comparison of the results for 
$g=1.5$ and $2$: $\tau_L$~(a) and $\tau_R$~(b) are shown for the both couplings at $n=14$ and $15$.}
\label{fig10}
\end{figure}
{\color{black} Finally, we compare the $t_A$-dependent relative enhancements of surface superconductivity, $\tau_L$ and $\tau_R$, for $g=1.5$ and $2$ in $n=14$ and $15$ chains, as shown in Fig.~\ref{fig10}. Data for $g=1.5$ are represented by curves with symbols, while data for $g=2$ are shown as solid curves. In panel (a), we find that the $\tau_L$ curves for both $g=1.5$ and $g=2$ have similar profiles and show no $n$-parity dependence. However, the enhancement $\tau_L$ is significantly greater for $g=1.5$ than for $g=2$.
In contrast, the behavior of $\tau_R$ in Fig.~\ref{fig10}(b) for $g=1.5$ exhibits strong $n$-parity effects. For a given $n$, the $t_A$-dependence of $\tau_R$ at $g=1.5$ is similar to its counterpart at $g=2$. Interestingly, the enhancement $\tau_R$ for $n=14$ at $g=1.5$ is greater than that at $g=2$, while the opposite is true for $n=15$. Therefore, we conclude that the characteristic features of the three surface critical temperatures and the $n$-parity effect of the right-end superconducting condensate persist at a smaller coupling strength $g$.}

\section{Conclusions}~\label{IV}

By numerically solving the self-consistent Bogoliubov-de Gennes equations, we have systematically investigated the end superconductivity in the $s-$wave superconducting Fibonacci chains $S_n$ with $n=11,\,...,\,16$, at half-filling. We find that the pair potential can be significantly enhanced at the ends of the chain by tuning the quasiperiodic strength $t_A/t_B$. This enhancement arises from the complex interplay between topological bound states and a constructive interference-like contribution from critical quasiparticles at the chain ends. Due to asymmetric configurations at opposite ends of Fibonacci chains, we observe different enhancement magnitudes at the left and right ends. Consequently, the system exhibits the three distinct superconducting temperatures: the left-end ($T_{cL}$), the right-end ($T_{cR}$), and the bulk ($T_{cb}$) critical temperatures. In the regime of the end enhancement, the system first transitions to the normal state in the chain center, followed by the ends. 

Our numerical results show that $T_{cL}$ and $T_{cb}$ are independent of the chain sequence number $n$~(when ignoring the finite-size effect at small $n$). The relative enhancement of $T_{cL}$ with respect to $T_{cb}$, governed by $\tau_{L} = (T_{cL}-T_{cb})/T_{cb}$, reaches its maximum of about $33\%$ at $t_{A}/t_B \approx 0.9$. In the regions $t_A/t_B < 0.2$ and $t_A/t_B > 3$, $T_{cL}$ approaches $T_{cb}$. Conversely, $T_{cR}$ shows strong dependence on the parity of $n$. Its relative enhancement with respect to $T_{cb}$, i.e. $\tau_{R} = (T_{cR}-T_{cb})/T_{cb}$, can be notably larger than $\tau_{L}$, reaching $\tau_{R} = 65$-$68\%$ for odd $n$ at $t_A/t_B = 1.6$. However, when $t_A/t_B \to 0$~(for odd $n$), $\tau_{R}$ becomes negative -- here, the right end transition to the normal state occurs before the chain center. For even $n$, $T_{cR}$ behaves similarly to $T_{cL}$, with noticeable differences only occurring near $t_{A}/t_{B}\approx 1.8$.

Additionally, based on this work and results given in Fig. 2 of  Ref.~\onlinecite{sun2024}, we anticipate significant oscillations in the end critical temperatures as functions of system size for small pairing couplings and short chain lengths. These finite-size effects fall outside the scope of our current study, as we employ sufficiently large chains to suppress them effectively.

Thus, our study unveils the complex superconducting order in quasicrystals, characterized by three distinct critical temperatures within a Fibonacci chain. {\color{black} A promising direction for future work would be a systematic statistical study of the surface critical temperatures across all possible segments of the same length obtained from the complete Fibonacci chain $S_n$}. Our findings are particularly relevant for research on quasicrystal physics and the design of novel superconducting materials.

\begin{acknowledgments}
A.A.S. acknowledges support from the International Academic Cooperation Programme of HSE University.
\end{acknowledgments}


\end{document}